# A Unified Deep Neural Network Potential Capable of Predicting Thermal Conductivity of Silicon in Different Phases

Ruiyang Li [a], Eungkyu Lee [a, *], Tengfei Luo [a, b, c, **]

[a] *Department of Aerospace and Mechanical Engineering, University of Notre Dame, Notre Dame, Indiana 46556, USA*

[b] *Department of Chemical and Biomolecular Engineering, University of Notre Dame, Notre Dame, Indiana 46556, USA*

[c] *Center for Sustainable Energy of Notre Dame (ND Energy), University of Notre Dame, Notre Dame, Indiana 46556, USA*

\* Corresponding author.

\*\* Corresponding author.

*E-mail addresses*: elee18@nd.edu (E. Lee), tluo@nd.edu (T. Luo).

**Abstract**

Molecular dynamics simulations have been extensively used to predict thermal properties, but simulating different phases with similar precision using a unified force field is often difficult, due to the lack of accurate and transferrable interatomistic potential fields. As a result, this issue has become a major barrier to predicting the phase change of materials and their transport properties with atomistic-level modeling techniques. Recently, machine learning based algorithms have emerged as promising tools to develop accurate potentials for molecular dynamics simulations. In this work, we approach the problem of predicting the thermal conductivity of silicon in different phases by performing molecular dynamics simulations with a deep neural network potential. This neural network potential is trained with *ab-initio* data of silicon in the crystalline, liquid and amorphous phases. The accuracy of our potential is first validated through reproducing the atomistic structures during the phase transition, where other empirical potentials usually fail. The thermal conductivity of different phases is then calculated, showing a good agreement with the experimental results and *ab-initio* calculation results. Our work shows that a unified neural network-based potential can be a promising tool for studying phase change and thermal transport of materials with high accuracy.

Predicting the thermal properties of materials in different phases is essential for material development for a wide range of applications. While temperature-dependent thermal properties of crystalline solids are necessary for many high temperature applications, it is critical to understand those of materials in the disordered phases as well. For instance, the thermal properties of amorphous materials could strongly affect the performance of semiconductor devices, such as photovoltaic solar cells and thin-film transistors [1,2]. Accurate estimation of the thermal properties in amorphous materials is very important for optimizing the thermal design of these devices. In addition, as single crystals are usually grown from the melts [3], the knowledge of the thermal conductivity in the liquid state is also of considerable importance.

Prediction of thermal properties, such as the thermal conductivity, is closely related to the microscopic insights into the vibrational dynamics of atoms, which can be simulated using molecular modeling. As a powerful modeling tool, molecular dynamics (MD) simulation has been extensively used to predict material properties and structures, but its accuracy depends on the model for interatomistic interactions. The *ab-initio* MD (AIMD) [4,5], while possessing the accuracy of density functional theory (DFT) calculations [6], is usually limited to small system sizes and time scales due to its high computational cost [7]. Modeling with larger structures and longer time is accessible only with empirical interatomistic potentials (EIP) in MD simulations, but there is limited guarantee for their accuracy and transferability. Even for common materials like crystalline silicon, EIPs fail to reproduce the phonon dispersion and thermal conductivity [8-10]. Better description of thermal transport in crystals can be achieved by resorting to the Boltzmann transport equation (BTE) method with the scattering rates predicted from the Fermi's Golden rule leveraging first-principle force constants [11-13]. However, this method does not work for disordered phases [14,15], which have complicated structures and a coexistence of localized vibrations and propagating modes [16]. Although MD simulations can naturally incorporate the influence of atomistic disorders, as outlined above, the accuracy of EIPs for describing the anharmonic effect is often questionable due to the simple functional form. Furthermore, it is noted that EIPs can be improved by physically and chemically motivated modifications, such as optimization for phonon properties [10]. But the situation can be dramatically different when dealing with materials in the liquid or amorphous phases, which have been proved rather difficult to model with a single EIP, as there is difficulty in describing different phases with the similar precision.

In the past few years, machine learning (ML) based techniques have been employed to harness the efficiency of EIPs while reconstructing the *ab-initio* potential energy surface. ML models like artificial neural networks (NN) [17], Gaussian process regression (GPR) [18] and others [19], are used as fitting tools to develop accurate interatomic potentials which can reproduce a certain amount of reference data generated from *ab-initio* calculations. During the training process, unknown parameters in the model are determined by minimizing the error between the given *ab-initio* data and prediction from the ML potential. After such a data-driven process, ML potential can precisely map the atomistic configurations to the corresponding energies and forces from the *ab-initio* data. Due to its ability of accurately reproducing *ab-initio* data, several different forms of ML potentials have been proposed to obtain the thermal and mechanical properties of crystalline solids such as Si [17], Zr [20], graphene [21] and single-layer $MoS_2$ [22]. Recently, it has been proved that more complex atomistic structures, such as amorphous GeTe [23] and polymer materials [24], can be simulated with similar methods as well. However, the capability of ML potentials in predicting thermal conductivity of a single material in different phases has not been evaluated.

With the power of ML techniques, it is possible to develop an interatomic potential such that we can predict the thermal conductivity of a material in different phases. As a testing material, silicon remains important to both solid state physics and semiconductor industry. It is also an ideal model material for such a study on the ML potential, due to a large number of EIPs and experimental results reported before. Moreover, accurate predictions of the thermal properties of multi-phase silicon with ML potentials can provide incentive for the accurate simulation of many other semiconductor materials. It will also provide a foundation for exploring thermal transport physics using MD simulations. In this study, we develop a unified neural network potential (NNP) to describe silicon in different phases and calculate their thermal conductivity values through lattice dynamics and MD simulations. This NNP is trained with *ab-initio* data of silicon in the crystalline, liquid and amorphous phases. We validate this potential by reproducing the atomistic structures during the phase transition, where other empirical potentials usually fail. The thermal conductivity of different phases is then calculated, showing a good agreement with the experimental results and *ab-initio* calculation results. Our work shows that a unified neural network-based potential can be a promising tool for studying phase change and thermal transport of materials with high accuracy.

In order to build the force field, we turn to the Smooth Edition of Deep Potential (DeePot-SE) method developed by Zhang et. al. [25], which is an end-to-end deep neural network (DNN) based potential scheme for MD simulations. We select DeePot-SE because it can train a unified model for sub-systems of different structures and sizes. In general, DeePot-SE represents the potential energy of each atomistic configuration as a sum of atomistic energies, which can be determined from the local environment of each atom. In DeePot-SE, a local coordinate frame is built for every atom and its neighboring atoms within the cutoff so that all the natural symmetries can be preserved. Further details about this scheme and related models can be found in the literature [26,27].

Here we briefly present how our neural network potential (NNP) for Si is developed in the form of DeePot-SE. First, we generate a reference dataset as the input for the training process. Since the objective of this study is to obtain the thermal properties of Si in different phases using only one potential, the dataset should be constructed with the *ab-initio* data of Si in different states. Therefore, the reference dataset consists of information (potential energy, atomistic forces, coordinates, supercell lattice vectors) of different configurations from AIMD simulations, together with structures with randomly displaced atoms at 0 K. Static self-consistent DFT calculations at 0 K are necessary to train the NNP with harmonic and anharmonic force constants of crystalline Si, where the lattice constant is changed by 0.5%, 1% and 2% to include the impact of volumetric strains. Some atoms in these 4×4×3 supercells are randomly displaced away from the equilibrium positions with at most 0.04 Å in each direction. As for the AIMD simulations at finite temperatures, atomistic structures are collected from different molecular trajectories in the isobaric-isothermal (NPT) ensemble with temperature ranging from 50 to 3000 K at zero pressure. A total of 35000 snapshots of crystalline, liquid and amorphous Si configurations are generated by the QUICKSTEP algorithm implemented in the CP2K package [28,29]. We employ the Goedecker-Teter-Hutter (GTH) double-$\zeta$, single polarization (DZVP-MOLOPT-GTH) basis set and the GTH-PBE pseudopotential [30] to represent core electrons, with the density cutoff set to be 300 Ry. After this database generating process, a potential is trained with the DeePMD-kit package [26]. Such potential model is then used in LAMMPS [31] to perform MD simulations.

Before the discussion of thermal properties calculated from MD simulations, the accuracy of this NNP model can be demonstrated by the differences between NNP predictions and AIMD results. We compare the NNP-predicted energies and forces with the separate AIMD datasets for testing, each of which is composed of 1000 snapshots of each phase that are not included in the training dataset. It is found that our NNP can provide similar results compared to the reference data, and the root-mean-square error (RMSE) of the energy is 0.27, 3.77 and 1.94 meV/atom for crystalline, liquid and amorphous configurations, respectively, while the RMSEs of the forces are 0.046, 0.27 and 0.19 eV/Å for the three sub-systems. Figure 1 shows the comparison of AIMD energies and NNP predicted energies, which indicates that NNP can reproduce the *ab-initio* energies of diverse structures.

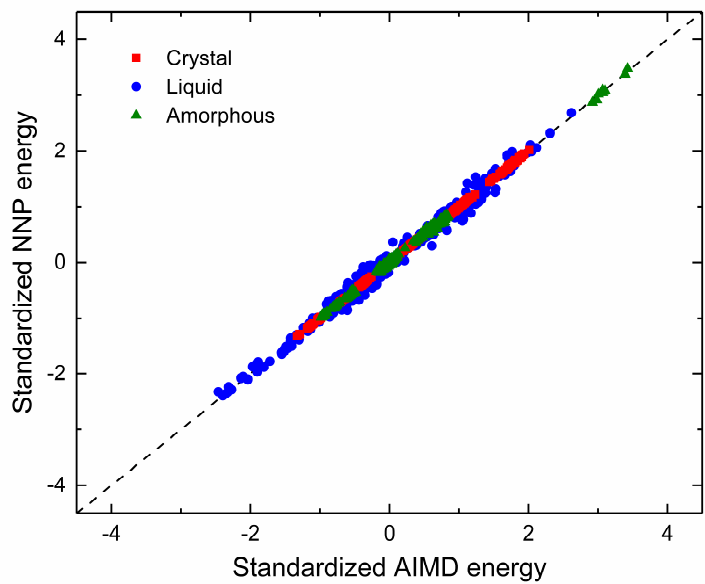

Figure 1. Standardized potential energies in different phases calculated using NNP compared with separate testing datasets from AIMD.

We first calculate the phonon dispersion of crystalline silicon (c-Si) using the DFT, NNP and EIPs, as shown in Fig. 2(a). It is observed that the dispersion from NNP shows excellent agreement with that from the DFT calculation, while the results using popular EIPs, such as Tersoff [32], Stillinger-Weber (SW) [33], and EDIP [34] potentials, obviously deviate from the DFT curve. Such agreement validates the capability to reproduce the second order force constants in c-Si with our NNP. The accuracy of third order force constants can be examined by the thermal conductivity calculation. The thermal conductivity of c-Si is estimated by the relaxation time approximation method implemented in Alamode [35]. Figure 2(b) shows the thermal conductivity as a function of temperature. Within the temperature range of 200 and 1000 K, the thermal conductivity values using NNP agree very well with DFT (<5% error) and experimental results [36,37].

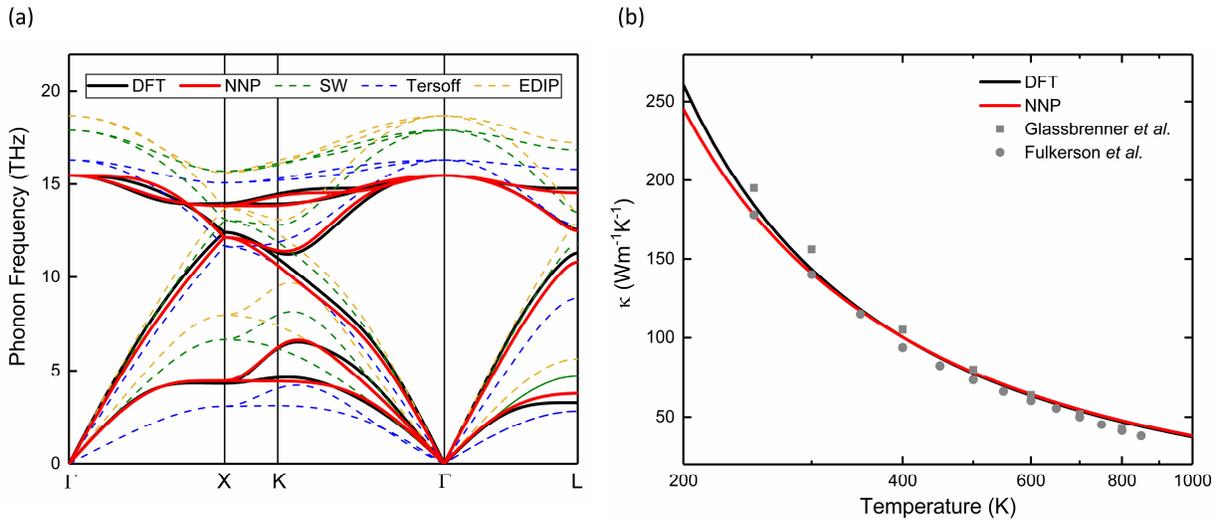

Figure 2. (a) DFT and NNP phonon dispersion of crystalline silicon, along with results from Tersoff, SW and EDIP empirical potentials. The second order force constants are well captured in NNP so that correct phonon dispersion can be obtained. (b) Thermal conductivity as a function of temperature for c-Si based upon DFT, NNP and experiments. NNP successfully produce thermal conductivity within 5% error of DFT results [36,37].

To predict the thermal conductivity of liquid and amorphous silicon, a precise reconstruction of the disordered morphologies is critical. Using NNP, we simulate liquid silicon (l-Si) in the NPT ensemble with pressure $P = 0$ GPa. A solid 4096-atom system is first heated from 0 to 3000 K over 50 ps, which serves as a quick melting process. Then this fully disordered structure is equilibrated at 1800 K for another 20 ps. To verify the liquid structure, we compare the radial distribution function (RDF) from simulations using NNP and EIPs, along with our AIMD results of a 384-atom structure. As shown in Fig. 3(a), there is very little difference in the RDFs from NNP and AIMD, and NNP shows a great reproduction of the first peak from AIMD at ~2.45 Å. However, empirical potentials like Tersoff and EDIP show significant deviation in RDF compared with the AIMD results, which means they cannot provide an accurate description of the liquid phase morphology. The only EIP with acceptable accuracy is SW, but it is still unable to reproduce the position of the first peak in RDF. The failure of EIPs in simulating l-Si may be attributed to their analytical functional forms. For example, Tersoff considerably overestimates the melting point, thus it is possible that the structures simulated with Tersoff at 1800 K are actually subcooled liquids. To further verify the accuracy of NNP in simulating solid-liquid transition at zero pressure, we determine the melting point by simulating a coexisting crystalline-liquid phase. The melting point with NNP is found to be $1510 \pm 10$ K at zero pressure, which is close to the DFT prediction (~1492 K with GGA) [38], but both are lower than the experimental value (~1685 K) [39]. It is known that GGA/PBE tends to overestimate the lattice constant [40] and thus underestimate the bonding strength. Therefore, it is understandable that DFT predicts a lower melting point than experiments. Since our NNP is trained on DFT data, the agreement with DFT melting point indicates that the NNP can accurately simulate the solid-liquid phase transition at zero pressure besides properly describing each phase.

The melt-quench procedure has been extensively employed to generate amorphous configurations in MD simulations [15]. In general, one starts with a well-equilibrated liquid and gradually lowers the temperature until the structure turns into an amorphous solid state. For the quench process, the cooling rate should be carefully set to obtain high-quality amorphous structures. However, for the quenching of a large size system, decent cooling rate values (~$10^{11}$ – $10^{12}$ K/s) are not accessible with the AIMD simulation due to its high computation cost [41,42]. We therefore use NNP to perform the quenching of a l-Si system with the cooling rate of $10^{12}$ K/s. The liquid is cooled from 1800 to 300 K in the NPT ensemble at zero pressure, and then relaxed into local energy minima. With this reasonable quench rate, we are able to generate reliable a-Si structures. To evaluate the quality of the final a-Si, we again compare the RDFs based on different methods, which is shown in Fig. 3(b). Compared with the RDF results of the a-Si structure derived from AIMD simulations, NNP successfully captures the first neighbor peak at ~2.35 Å and another broad peak at ~3.84 Å, with no other features between these two peaks, indicating that NNP can generate high-quality a-Si. However, even though EDIP does a decent job in describing the a-Si structure, RDFs obtained by Tersoff and SW show large difference from the AIMD results, especially at the valley region between the first two peaks. Even for EDIP, there is an unexpected small bump between the first two peaks, and the first peak is overestimated compared to AIMD results. Once we obtain the good l-Si and a-Si structures in MD, we can proceed to calculate their thermal properties. In this study, we use Green-Kubo (GK) method [43] to predict the thermal conductivity of these two disordered Si states.

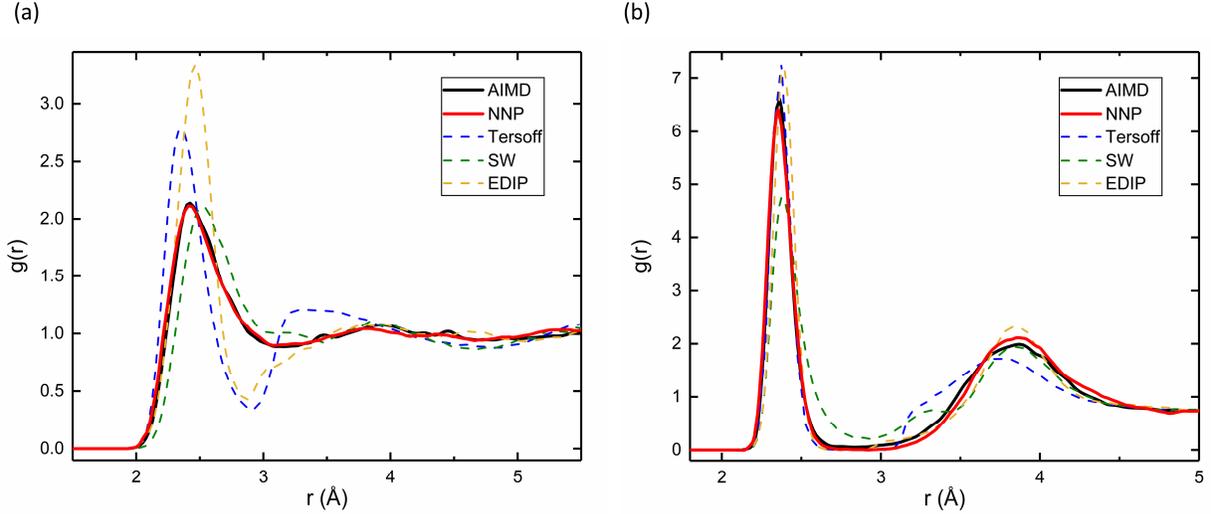

Figure 3. Radial distribution functions for (a) liquid silicon (l-Si) and (b) amorphous silicon (a-Si). Both are 4096-atom systems generated at zero pressure, with equilibration for l-Si at 1800 K and a-Si at 300 K. The red solid lines indicate NNP results, and dash lines in various colors indicate results using other empirical potentials. The AIMD results of a 384-atom structure are shown as the black solid lines for comparison.

In the GK method for an isotropic system, the thermal conductivity is related to the heat current autocorrelation function (HCACF) via the Green-Kubo formula,

$$\kappa = \frac{V}{3k_B T^2} \sum_{\alpha=1}^{3} \int_0^\infty \langle J_\alpha(0) J_\alpha(t) \rangle dt, \quad (1)$$

where $V$, $\alpha$, $J_\alpha(0)$ stand for the volume of simulation box, the $\alpha$-th coordinate direction and the $\alpha$-th component of the heat current, respectively. The HCACF can be obtained from equilibrium MD (EMD) simulations, in which the heat current is evaluated as

$$\boldsymbol{J} = \frac{1}{V} \sum_i (E_i \boldsymbol{v}_i - \boldsymbol{S}_i \cdot \boldsymbol{v}_i), \quad (2)$$

where $E_i$ and $\mathbf{v}_i$ denote the energy and velocity of the atom $i$, and the per-atom virial stress $\mathbf{S}_i$ is expressed as the outer product of relative position $\mathbf{r}_{ij} = \mathbf{r}_j - \mathbf{r}_i$ and derivative of potential energy of the neighboring atom $U_j$ with respect to the relative position:

$$\mathbf{S}_i = \sum_{j \neq i} \mathbf{r}_{ij} \otimes \frac{\partial U_j}{\partial \mathbf{r}_{ji}}. \tag{3}$$

Specifically, after the relaxation process with NPT ensemble we perform 2 ns-long simulations in the microcanonical (NVE) ensemble to collect the heat current data. Using GK method, we estimate the lattice thermal conductivity of Si near the melting point and that of amorphous structures at 100 – 300 K. For each configuration of over 20000 atoms, we average the thermal conductivity values from 8 independent runs with different initial atomistic velocity distributions. Figure 4(a) shows the averaged integration of the thermal conductivity of c-Si at 1200 K, and the inset is a plot for the representative normalized HCACF. It is observed that a correlation time of 100 ps is sufficient to achieve a converged $\kappa$ value of the solid structure.

In general, thermal conductivity is expressed as the sum of lattice (phonon) thermal conductivity $\kappa_l$ and electronic thermal conductivity $\kappa_e$ as $\kappa = \kappa_l + \kappa_e$. For c-Si below 1000 K, phonon is the dominant heat carrier [37]. Above 1000 K, contributions from the electrons and holes, which consist of an electronic polar part and an ambipolar part [37], are not negligible and must be taken into consideration. The overall $\kappa$ of c-Si above 1000 K can be determined by adding up the theoretical values of $\kappa_e$ and the calculated $\kappa_l$ based on the GK method. When the temperature is increased over the melting point, silicon turns into a metallic liquid with the persistence of covalent bonding [42], and the major contribution to the heat conduction is from the electrons and holes. Here we apply the Wiedemann-Franz Law to determine $\kappa_e$ in the melt state, which can be written as

$$\kappa_e = LT\sigma, \tag{4}$$

where $L$ is the theoretical Lorenz number for free electrons (2.45×10$^{-8}$ W$\Omega$K$^{-2}$), and $\sigma$ is electrical conductivity. For the value of $\sigma$ for l-Si, we neglect the weak temperature dependence reported in the literatures and use the experimental value of $\sigma = 1.36 \pm 0.03$ μΩ$^{-1}$m$^{-1}$ [44]. The temperature-dependent $\kappa$ of Si above 1000 K is shown in Fig. 4(b). For the data point at 1600 K, we use the same $\sigma$ as the one at 1700 K given its reported weak temperature dependence. In Fig. 4(b), our predicted $\kappa$ values below the melting point, which include the electron and phonon contributions, do not differ much from the experimental results [37,45]. The lower $\kappa$ using NNP may be a result from the lower melting temperature we have predicted (1510 ± 10 K), in which case $\kappa$ drops more quickly as the temperature approaches the melting point. For l-Si, the phonon thermal conductivity is only about 0.7 ± 0.08 Wm$^{-1}$K$^{-1}$, with no temperature dependence in the range of 1600 – 1800 K. When electron contributions are included, an abrupt increase in the total thermal conductivity is observed just across the melting point. It is noted that since the phonon contribution is small, the uncertainty of our predicted $\kappa$ of l-Si should be attributed to the Wiedemann-Franz relation, where the Lorenz number is derived based on the free electron model. In general, it is clear that with NNP we can obtain $\kappa$ values in reasonable agreement with the measurements below and above the melting point.

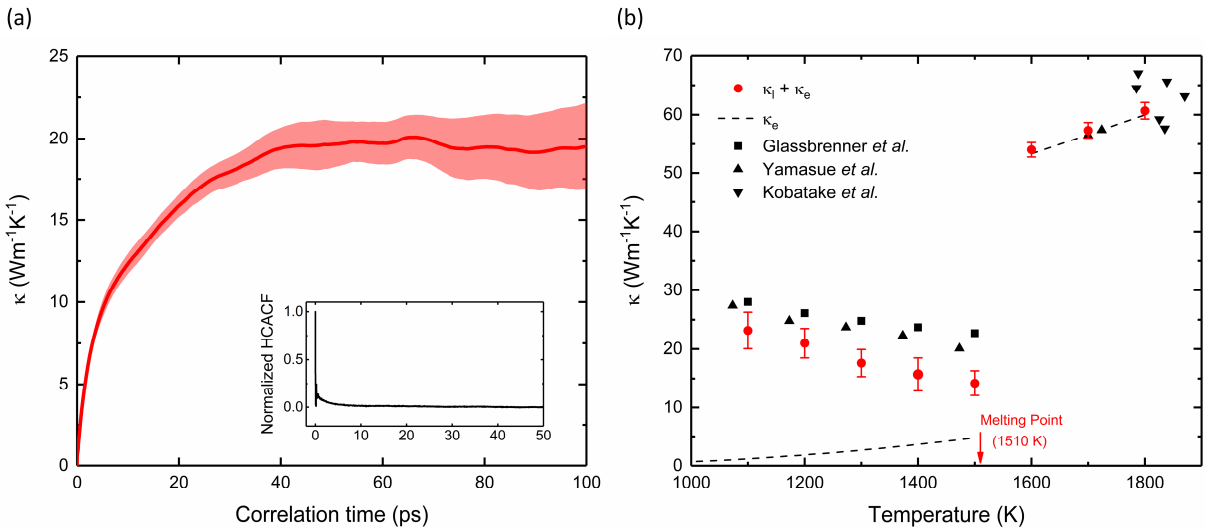

Figure 4. (a) Thermal conductivity of c-Si at 1200 K using the GK method, and the shaded area indicates the standard deviation among 8 independent simulations. The inset shows a representative normalized HCACF. (b) Thermal conductivity of Si above 1000 K from MD simulations (red), and literature experiments (black symbols [37,45,46]). Electronic thermal conductivities $\kappa_e$ (black dash line) have been included in the calculation. The contribution of electrons is dominant for l-Si, with the lattice thermal conductivity $\kappa_l$ = 0.7 ± 0.08 Wm$^{-1}$K$^{-1}$. The melting point predicted by NNP is 1510 ± 10 K.

For a-Si, the calculated temperature-dependent lattice thermal conductivity using GK method with NNP are plotted in Fig. 5(a). The thermal conductivity of a-Si increases as the temperature increases from 100 to 300 K. We focus on this temperature range because the experimental results are only available at up to ~300 K. It is found that $\kappa$ estimated by NNP are similar to the experimental values. Specifically, our $\kappa$ value at 300 K is 1.22 ± 0.16 Wm$^{-1}$K$^{-1}$, which is within the range (1–2 Wm$^{-1}$K$^{-1}$) of the measured $\kappa$ of a-Si thin films in most studies [47-51]. This value is also close to the thermal conductivity contributed by the non-propagating vibrational modes (diffusons) predicted by the Allen-Feldman theory (1.2 ± 0.1 Wm$^{-1}$K$^{-1}$) [52]. We note that many MD and theoretical studies show various results ranging from 1 to 3 Wm$^{-1}$K$^{-1}$ [52-56], depending on the different EIPs and system sizes used. Moreover, recent experimental works show large variations due to different material preparation and measurement techniques, and the measured $\kappa$ of thicker a-Si films can be as high as 4 Wm$^{-1}$K$^{-1}$ at 300 K [57,58], with no convergence as the film thickness increases. Our lower $\kappa$ can be attributed to the limited size of a-Si system in MD simulations [54,59], which suppresses the contribution from propagating modes (propagons) [16]. Recent simulation studies found that propagons could contribute significantly, up to 40% or a half, of the total $\kappa$ of a-Si [15]. It has also been proved by TDTR measurement that a-Si films below 100 nm show a relatively constant $\kappa$ (~1.1 Wm$^{-1}$K$^{-1}$) that is dominated by diffuson transport [58], which is consistent with our prediction with the system length of ~20 nm. In addition to the thermal conductivity calculation, the phonon density of states of a-Si at 300 K is shown in Fig. 5(b) together with the experimental results [60], serving as an indication that NNP provides a reasonably good description of vibrations in a-Si.

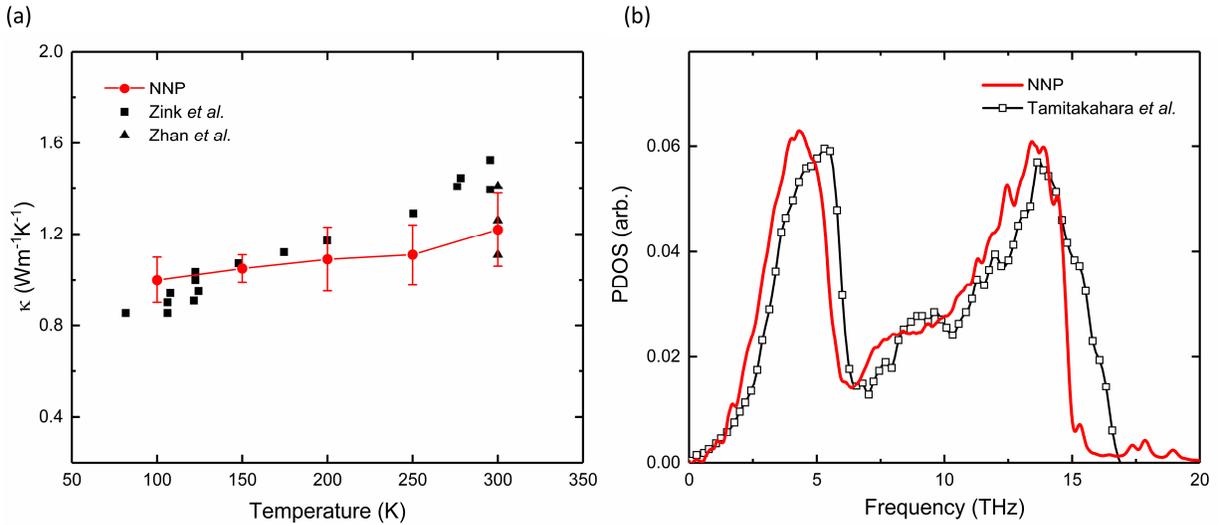

Figure 5. (a) Calculated thermal conductivity as a function of temperature for amorphous configurations with NNP (red). Black symbols indicate the thermal conductivities by measurements [51,61]. (b) Phonon density of states (PDOS) for a-Si at 300 K using NNP, compared with experimental results [60].

In conclusion, a unified neural network potential (NNP) is developed for prediction of the thermal conductivity of Si in the crystalline, liquid and amorphous phase. Using MD simulations with NNP, we accurately reproduce the atomistic configurations of Si in comparison to the AIMD results in different phases and also during the phase transition, with the predicted melting point close to the DFT result. Especially for the disordered morphologies of Si, such as liquid and amorphous states, NNP outperforms classical force fields in generating high-quality disordered structures. The phonon dispersion and thermal conductivity of c-Si with NNP are found to agree well with DFT results. It is also observed that our predicted thermal conductivity values around the melting point, which include the electron contributions, show a good agreement with the experimental values. Furthermore, NNP can describe the vibrations and thermal transport in a-Si, with the predicted thermal conductivity close to the one dominated by non-propagating vibrational modes. This study provides a solution for predicting thermal properties of multi-phase systems with high accuracy, showing that the NNP is a powerful tool for modeling thermal transport in distinct structures of materials.

**Acknowledgements**

The authors would like to thank ONR MURI (N00014-18-1-2429) for the financial support. The simulations are supported by the Notre Dame Center for Research Computing, and NSF through the eXtreme Science and Engineering Discovery Environment (XSEDE) computing resources provided by Texas Advanced Computing Center (TACC) Stampede II under grant number TG-CTS100078.